\documentstyle[preprint,aps]{revtex}

\begin{document}

\draft

\title{
\begin{flushright}
{\rm Oslo-TP-9-94, TP-MUBR 94-06/1\\
gr-qc/YYMMXXX}
\end{flushright}
 A reinterpretation of the Taub singularity}

\author{Bj\o rn Jensen \footnote{Electronic address: BJensen@anyon.uio.no}}

\address{Institute of Physics, University of Oslo,
P.O. Box 1048, N-0316 Blindern, Oslo 3, Norway}

\author{ Jarom\'{\i}r Ku\v{c}era \footnote{Electronic address:
JKucera@elanor.sci.muni.cs and JKucera@csbrmu11.bitnet}}

\address{Dep. of Theor. Physics and Astrophysics, Masaryk University\\
Kotl\'{a}\v{r}sk\'{a} 2, 611 37 Brno, Czech Republic}

\date{June 20, 1994}

\maketitle

\begin{abstract}
We reinterpret the well known Taub-singularity
in terms of a cylinder symmetric geometry. It is
shown that a cylindrical analog to the Einstein-Rosen
bridge as well as a cosmic string will be present in
the geometry.
\end{abstract}

\vspace{1.5cm}


\newpage

It is a somewhat curious fact that the static spherical symmetric vacuum
solution of Einstein's field equations
is unique while no unique solution exist for the corresponding
cylinder symmetric problem. Even though the most general form
of the cylinder symmetric vacuum solution, the Levi-Civita metric \cite{Levi},
 has been known for a long
time physical interpretations has only been achieved for a relatively
restricted class of solutions \cite{Bonnor}.
The aim of these notes is to provide a reinterpretation
of the well known Taub singularity \cite{Taub} in terms
of a cylinder symmetric geometry. Along the way we
give by this approach an interpretation of two
previously ill-understood solutions of the Levi-Civita metric.\\

\smallskip

The structure of these notes is as follows.
We will first use a well known connection between
plane and cylinder symmetry in order
to construct the space-time manifold
of a cylinder symmetric shell of finite thickness
which is filled with
an incompressible perfect fluid.
This connection has not,
as far as we know,
been utilized before in the study of cylinder
symmetric systems in the literature. A cylinderical analog
to the Taub-singularity resides in the vacuum
 on one side of the shell while
the other vacuum is flat.
These vacuum solutions have appeared in previous studies but
have so far defied interpretation \cite{Bonnor,Martins}.
We match this construction using the Israel-formalism
to the conic flat space which appear outside cosmic gauge strings.
The resulting structure can be
interpreted either as a cosmic gauge string
with a naked singularity in the interior region
or as a naked cylinder-symmetric singularity with a
cosmic string in the interior. It is
possible to travel from the conic region and into
the singular region after having been traveling through a flat part
which we interpret as a cylinderical analog to the Einstein-Rosen bridge.\\

\smallskip

In \cite{Avak,NKH} a static solution of Einsteins equations
was obtained which was interpreted as an infinite plane
wall of finite thickness composed of an incompressible ideal perfect fluid
with a constant energy density.
The pressure inside the wall is everywhere greater than zero and the boundaries
of the wall are derived from the physical condition
of vanishing pressure $p=0$. This particular solution has a strange and
unexpected property. Despite the fact that the
energy density is assumed to be constant inside the wall
the interior solution can in no way be made mirror symmetric about the central
plane in the wall if the solution is to be free from curvature
singularities. It follows that the exterior solution on
one side of the wall is flat while the exterior solution on the other side
must display a non-vanishing intrinsic curvature \cite{NKH}.
This property is in fact not a
curious property of the specific
solution \cite{NKH} only. It can be shown in general that
no singularity free plane symmetric source which obeys the
dominant energy condition will give rise to the same geometry
on both sides of the wall \cite{Dolgov}.
It was a wish to get a deeper understanding of this curious
property that led to the present notes.
We will first use this specific solution \cite{NKH} and a connection
between plane and cylinder symmetry in order to
derive the space-time manifold of a thick cylindrical shell.\\

\smallskip

The connection between plane and cylinder symmetry
which we want to utilize is a simple one.
Every static cylinder symmetric interval can be
written in the form \cite{HN}
\begin{equation}
\label{interval}
ds^{2}=e^{2 \nu(r)} dt^{2}-dr^{2}-e^{2 \chi(r)} d\varphi^{2} -e^{2 \lambda(r)}
dz^{2}.
\end{equation}
In these coordinates we will assume $0\leq \varphi \leq 2 \pi$ and we
will leave  the range of $r$ and $z$ unrestricted.
Similarly, every static plane symmetric interval can be brought on the form
\cite{HN}
\begin{equation}
\label{intervalpl}
ds^{2}=e^{2 \nu(x)} dt^{2}-dx^{2}- e^{2 \lambda(x)}(dy^{2} + dz^{2}).
\end{equation}
Hence, by imposing the equality $g_{\varphi \varphi}=g_{z z}$ in eq.
(\ref{interval}) we clearly see that the cylinder symmetric case is locally a
special case of the plane symmetric one. We will first discuss which
axisymmetric vacuum solutions are possible with this sort of restriction.\\

\smallskip

The most general axisymmetric vacuum solutions, known as the Kasner  solutions,
can in general be written in the
convenient form \cite{FREH1,FREH2}
\begin{equation}
\label{genkas}
ds^{2}= -\left({dr^2+r^{4\left(\frac{A+1}{A^2+3}\right)}
d\varphi^2+r^{2\left(
\frac{A^2-1}{A^2+3}\right)} dz^2 }\right)+r^{4\left(\frac{1-
A}{A^2+3}\right)}
dt^2.
\end{equation}
Here $A$ is an arbitrary constant with the limits $A
\rightarrow \pm
\infty$ included. Imposing the ``plane symmetry''
condition on
this metric
restricts the constant
$A$ to the two
possibilities $A=3$ and $A=-1$.
For $A=3$ we obtain the line-element
\begin{equation}
\label{aeq3}
ds^{2}= -\left({dr^2+r^{4/3} d\varphi^2+r^{4/3}
dz^2 }\right)
+r^{-2/3} dt^2
\end{equation}
which is regular everywhere except at $r=0$ where a
 curvature
singularity resides. It is clear
that this geometry can be considered as a
cylindrical analog to the Taub solution
with the Taub singularity residing
at $r=0$ \cite{Taub}. The metric structure does not
reduce
to the Minkowski metric
in cylinder coordinates in any region. This is a
well known
problem which arises in the interpretation
of cylinder
symmetric
solutions in general. However, all the curvature
invariants vanish in the limit $r\rightarrow\infty$.
For $A=-1$ the interval eq.(\ref{genkas}) reduces to
\begin{equation}
\label{aeq1}
ds^{2}= -\left({dr^2+d\varphi^2+dz^2 }\right)
+r^2 dt^2,
\end{equation}
which is regular and flat everywhere. However, a coordinate
singularity is present at $r=0$.\\

\smallskip

At this point it is
interesting to
make a connection to the form of the Levi-Civita metric used in
\cite{Bonnor}. In that work the Levi-Civita metric was written
as
\begin{equation}
ds^2=R^{8\sigma^2 -4\sigma}(K^2dR^2+L^2dz^2)+M^2R^{2-4\sigma}d\phi^2-
N^2R^{4\sigma}dt^2\,\,\,\, (R\geq 0)\, .
\end{equation}
$K,L,M,N$ and $\sigma$ was considered as arbitrary constants.
Imposing the plane symmetry condition implies that
\begin{equation}
\sigma=\pm\frac{1}{2}\, .
\end{equation}
In previous studies these solutions have proven particularly
difficult
to interperet \cite{Bonnor}.
$\sigma =1/2$ leads directly to eq.(5) in our work while
$\sigma =-1/2$ leads to eq.(4) via a coordinate transformation.
We therefore believe that the present considerations are of
particular
importance for the understanding of these solutions.\\

\smallskip

We next turn our attention to the metric field inside the cylindrical wall.
We assume a cylindrical wall of finite thickness and composed of a fluid with a
non-zero constant and positive energy density $\rho$.
The pressure is everywhere positive $p > 0$ except at the boundaries of the
wall where $p=0$ holds.
We will denote the radial coordinate in this region by $R$.
On the assumption that the metric structure takes the form
 eq.(\ref{interval}) we find that
Einsteins equations for the case of an incompressible
ideal perfect fluid, which is at rest
in the coordinate system, are given by
\begin{eqnarray}
\label{ein0}
G^t\,_t &=& \chi''+\lambda''+\lambda'^{2}+\chi'^{2}+\lambda' \chi'=-\rho
\\
\label{ein1}
G^r\, _r &=& \nu' \lambda'+\nu' \chi' +\lambda' \chi'=p
\\
\label{ein2}
G^\varphi\, _\varphi &=& \nu''+\lambda''+\nu'^{2}+\lambda'^{2}+\nu' \lambda'=p
\\
\label{ein3}
G^z\, _z &=& \nu''+\chi''+\nu'^{2}+\chi'^2+\nu' \chi'=p.
\end{eqnarray}
$'$ denotes differentiation with respect to the radial coordinate $R$.
We now impose the ``plane symmetry'' restriction $\lambda(R)=\chi(R)$.
Then the form of the metric, and hence Einsteins equations, for the cylinder
and plane
symmetrical situations coincide.
In \cite{NKH} the plane symmetric situation was described using Cartesian
coordinates $x,y,z$. The coordinates are chosen in such  a way that the vector
$\partial_z$ is always perpendicular to the plane wall while
$\partial_x$ and $\partial_y$ are parallel to it. Identifying the
$z$-coordinate in \cite{NKH} with $R$, the
$x$-coordinate with $\varphi$ and the $y$-coordinate with $z$ in
eq.(\ref{genkas}), we can easily transform the solution found in \cite{NKH} to
cylindrical coordinates. We then get
\begin{eqnarray}
\label{elda}
e^{2 \lambda(R)}&=&\cos^{(4/3)} \beta R
\\
\label{enu}
e^{2 \nu(R)}&=& \left[ {\frac{-Q \sin \beta R +F(-\frac{1}{2},-\frac{1}{6};
\frac{1}{2};\sin^{2} \beta R)}{\cos^{1/3} \beta R}}\right] ^{2}
\\
\label{press}
p(R)&=&\rho (e^{-\nu(R)}-1).
\end{eqnarray}
Here $\beta=\frac{1}{2} \sqrt{3\rho}$,
$F(a,b;c;x)$
denotes the hypergeometric function, and $Q$ is a constant of integration.
A detailed discussion of this solution allow us to restrict our attention to
situations with $0<Q<F(-\frac{1}{2},-\frac{1}{6};\frac{1}{2};1)$ and
$0\leq R<\pi/(2\beta )$
(due to periodicity of the trigonometric functions in
(eq.\ref{elda}-\ref{press})) \cite{NKH}.
However, in order not to violate the dominant energy condition
$Q$ should be further restricted to the interval
$0<Q<Q_m$ where $Q_m\sim 0.86$
(the upper limit can only be determined numerically).
Then two kinds of boundary conditions,
 both characterized by $p=0$, follow from eq.
(\ref{ein1}) which now takes the form
\begin{equation}
\label{twokind}
\lambda' (2 \nu'+ \lambda')=0.
\end{equation}
It follows that
one kind of boundary is characterized by
\begin{equation}
\label{firstb}
\lambda'=0,
\end{equation}
while a second one is given by
\begin{equation}
\label{secondb}
\lambda'=-2\nu'.
\end{equation}
When eq.(16) is combined with the condition $0\leq R<\pi /(2\beta )$ it
follows that $R=0$. Equation (17) is much harder to solve but it is
readily shown numerically that it possess one solution such that
$R<\pi /(2\beta )$ \cite{NKH}.\\

\smallskip

We now want to match this interior solution smoothly
to the two axisymmetric vacuum solutions considered earlier. It is seen from
the conditions of continuity
of the metric tensor and its first derivatives that
only metric eq.(\ref{aeq1}) is possible to join to the boundary of the first
kind eq.(\ref{firstb}) while only metric eq.(\ref{aeq3}) is
possible to join to the boundary of the second kind eq.(\ref{secondb}).
Since the metric and its derivatives are continuous across the
matching surfaces it should be possible to set up a single
coordinate system for the entire manifold. It is rather
straightforward to extend the $R$-coordinate into the flat interior
part via the coordinate transformation $r=Q\beta (R-1/(Q\beta ))$,
 $t\rightarrow Q\beta t$.
The metric eq.(5) then takes the form
\begin{equation}
\label{vnitrni}
ds^{2}= -\left({dR^2+d\varphi^2+dz^2 }\right)
+Q^2 \beta^2 [R - 1/(Q \beta)]^2 dt^2,
\end{equation}
where now $R\leq 0$.
The metric component $g_{\varphi \varphi}$ in eq.(\ref{vnitrni})
is independent of
$R$. Hence a space-like hyper-surface
 \, $t=\mbox{const.}$, $z=\mbox{const.}$\, has a cylindrical geometry
with a
$\mbox{S}^1 \times
{\rm I \hspace{-.6mm} R}$
product topology.
It is also possible to extend the coordinatization into the
Taub-part of the manifold. However,
contrary to the claim in \cite{NKH}
it is not possible to match the wall smoothly to the metric eq.(4)
in such a way that the singularity disappears, i.e. it is
not possible to replace the Taub-singularity with the
plane wall solution in \cite{NKH}.
This follows immediately from the condition for the
continuity of the gravitational potential $g_{tt}$ and
the corresponding derivative relation across the
matching surface. However the solutions
can be matched smoothly when the presence of the singularity structure
is tolerated (this must partially be done numerically
due to the complex form of the metric coefficients in the interior
of the wall). The singularity will then reside at a finite
distance from the surface of the wall.
It follows that the plane solution found
in \cite{NKH} is the first non-singular source found
for the Taub-singularity. Other previous known sources
for the Taub-singularity display infinitely
thin walls with and without a scalar field as part of the
vacuum structure \cite{Nov,Ipser,Gron}.\\

\smallskip

We will now match the above structure
consisting of the plane and the two
vacuum solutions to the exterior region
of a cosmic gauge string. The geometry in this region is
assumed to take the form
\begin{equation}
ds^2=Fdt^2-B^2(R+R_0)^2d\phi^2-dz^2-dR^2\,\,\, (R<0)\, .
\end{equation}
Here $F$, $B$ and $R_0$ are constants to be determined.
When $B\neq 1$ this geometry displays a conic deficit angle
which is a well known feature of the cosmic gauge string
exterior space-time \cite{Gott}.
The matching will require that a thin cylindrical shell $S$
in general will be induced. Let $\vec{N}$ be a space-like
unit vector orthogonal to this surface and we will let
it point from the cosmic string geometry and into
the region described by
eq.(18). We will let $S$ reside at a fixed
radial coordinate $R=R_S<0$.
The properties of $S$ can be
computed from the extrinsic curvatures $K_{ij}$ of this surface
relative to the two flat geometries by the use of
the Israel-formalism \cite{Israel}. When $S_{ij}$ denotes the energy-momentum
tensor of the matter in $S$ we have
\begin{equation}
-S_{ij}=\gamma_{ij}-g_{ij}\mbox{Tr}\gamma_{ij}
\end{equation}
where $\gamma_{ij}=K^{+}_{ij}-K^{-}_{ij}$. $K^{+}_{ij}$
is the curvature relative to the flat metric eq.(18) while
$K^{-}_{ij}$ is the extrinsic curvature of $S$ relative to the
gauge string metric eq.(19).
{}From the condition for the continuity of the gravitational potentials
at $R=R_S$
we get $F=Q^2\beta^2(R_S-1/(Q\beta ))^2$ and $B=(R_S+R_0)^{-1}$.
{}From the definition $K_{ij}=-N_{i;j}$
we then have
\begin{eqnarray}
S_{tt}&=&\frac{Q^2\beta^2 (R_S-1/(Q\beta ))^2}{R_S+R_0}\\
S_{\phi\phi}&=& \frac{1}{R_S-1/(Q\beta )}\\
S_{zz}&=&\frac{1}{R_S-1/(Q\beta )}-\frac{1}{R_S+R_0}\, .
\end{eqnarray}

 The metric eq.(19)
is trivially regular provided $R+R_0<0$. However this
breaks both the
weak and the dominant energy-conditions since
$S_{\hat{t}\hat{t}}\leq 0$
(hatted indices refer to the tetrade observer
at rest in the geometry).
The strong energy condition is not
satisfied either since the Tolman-mass density
$S^{\hat{t}}\, _{\hat{t}}-S^{\hat{z}}\, _{\hat{z}}-
S^{\hat{\phi}}\, _{\hat{\phi}}=2Q\beta (Q\beta R_S-1)^{-1}$
is negative.
This is unexpected since the intrinsic curvatures
in the cosmic gauge string metric vanish.
The geometry eq.(19) is boost-invariant
in the longitudinal direction. It is
worth noting that this symmetry is not
reflected in the energy-momentum tensor of the matter
in $S$. The surface $S$ is a cylindrical surface and
in what way it ``bends'' can be found quite easely from the
sign of $K_{\phi\phi}$. Relative to the metric eq.(19)
we have $K_{\phi\phi}=(R_S+R_0)^{-1}$. Since $K_{\phi\phi}<0$
when $R_S+R_0<0$ we have that $\vec{N}$ can be looked upon as
pointing ``into''
a cylinder defined by $S$. In this situation we can interperet
the resulting structure as representing a cosmic string with a
naked singularity in the interior region. The weak energy
condition is satisfied when $R_0$ takes on sufficiently
positive values. In this situation eq.(19) is no longer
regular but will display a conic singularity at $R=-R_0$ which
we take as an indication for the presence of a cosmic string.
Such singularities are very weak in the sense that the singular
axis can be replaced with a smooth solution of Einsteins
equations \cite{Hiscock}. We now have that $K_{\phi\phi}>0$
and we can look upon $\vec{N}$ as pointing ``out of'' a
cylinder defined by $S$. In this situation $S$ actually
``encapsulates'' a cosmic string such that the
 string defines a symmetry axis in the manifold.\\

\smallskip

It is interesting to
investigate the motion of freely falling particles moving
in the radial direction
in the flat region eq.(18).
In this geometry we get
from the geodesic equation that the radial acceleration
of the particle relative
to a coordinate-basis observer is
\begin{equation}
\ddot{R}=\frac{-E^2}{Q\beta (|R|+1/(Q\beta ))}\, .
\end{equation}
$\ddot{}$ denotes twice differentiation with respect to
an affine parameter and
the conserved energy of the particle $E$ obeys $E=Q^2\beta^2(R-
1/(Q\beta ))^2\dot{t}$. It is clear that the particles experiences
a positive acceleration and will consequently
experience the shell with the
perfect fluid as source for attractive gravitation.
This surface displays a positive gravitational mass
and is therefore a natural
source for an attractive gravitational pull.
This fact explains the negative Tolman-mass density
of the shell $S$ which screens the field of the wall.
Note that the acceleration is everywhere finite.
Hence nothing seems to prevent a freely falling particle
to enter the wall and subsequently to get in
``direct'' contact with the singularity.
In the ``Taub-part'' of the manifold the sign of $\ddot{R}$
changes. Hence the particle will experience the singularity
as source for a repulsive gravitational ``force''. The
repulsive nature of the Taub-singularity is in accord with
the repulsive nature of other well known singularities
such as the singularities in the Reisner-Nordstr\"{o}m and
Kerr-geometries. However the Taub-singularity differs
from these since it is not hidden behind any event-horizon.\\

\smallskip

The Taub singularity challenges our understanding of the
singularity structure in the general theory of relativity.
When interpreted as a ``sheet'' singularity it can obviously
be generated by a source with a {\it finite} gravitating
energy density (which obeys the energy conditions) and
without the occurrence of topology changing or causality
violating surfaces or regions. It is usually believed
that singularities in Einsteins theory is connected with
such ``pathologies'' or
(near) infinite energy densities \cite{Joshi}.
Due to the fact that no physically acceptable sources
(i.e. sources that obey the dominant energy condition)
will generate the same geometry on both sides of a wall and
to conform with
the above belief we are therefore tempted to interpret the Taub
solution as describing a cylindrical ``sheet'' singularity and
not a ``planar'' sheet singularity as is
the usual interpretation \cite{Ipser}
when it is generated by physically
acceptable sources. The existence of the
Taub-singularity is in this interpretation directly
coupled to the existence of a topology changing region
described by eq.(18).\\

\smallskip

{\large \bf Acknowledgments}
\\
One of us (J.K.) thanks the University of Oslo for hospitality
during the time which parts of this work was performed and
acknowledges the Norwegian Research Council NAVF/SEP for financial support.
We also thanks Svend Hjelmeland, a master student at UiO, for
interesting conversations concerning the
properties of the Levi-Civita metric \footnote{When these
notes was at the end of completion we received a preprint
gr-qc/9405074 where it also is shown that
Einstein-Rosen like bridges may exist ``inside'' cosmic
gauge strings.}.

\end{document}